\documentclass[final]{aipproc}

\layoutstyle{6x9}

\usepackage{amsmath,amssymb,amsfonts,amsthm}
\usepackage{hyperref}
\usepackage[mathscr]{eucal}

\newtheorem*{thm}{Theorem}

\theoremstyle{definition}
\newtheorem{de}{Definition}
\newtheorem*{exa}{Example}
\newtheorem*{exas}{Examples}

\def\co{\colon\thinspace}

\renewcommand{\leq}{\leqslant}
\renewcommand{\geq}{\geqslant}

\DeclareMathOperator{\ev}{ev}

\DeclareMathOperator{\Ber}{Ber}

\DeclareMathOperator{\tr}{tr} 

\DeclareMathOperator{\str}{str}

\DeclareMathOperator{\ber}{ber}

\DeclareMathOperator{\Sym}{Sym}

\newcommand{\wed}{{\Lambda}}

\newcommand{\RR}{\mathbb R}

\newcommand{\ZZ}{{\mathbb Z}}

\newcommand{\CC}{\mathbb C}
\newcommand{\NN}{{\mathbb N}}

\newcommand{\f}{\mathbf{f}}

\newcommand{\g}{\mathbf{g}}

\def\a{\alpha}

\newcommand{\F}{{\Phi}}

\newcommand{\xp}{{\mathbf{x}}}

\newcommand{\ps}{{{\psi}}}

\begin{document}

\title{Operators on superspaces and generalizations of the
Gelfand--Kolmogorov theorem}


\author{H.~M.~Khudaverdian}{
  address={School of Mathematics, The University of Manchester, Oxford Road, Manchester, M13 9PL, UK} }

\author{Th.~Th.~Voronov}{
  address={School of Mathematics, The University of Manchester, Oxford Road, Manchester, M13 9PL, UK}
}

\begin{abstract}
Gelfand and Kolmogorov in 1939 proved that a compact Hausdorff
topological  space $X$ can be canonically embedded into the
infinite-dimensional vector space $C(X)^* $, the dual space of the
algebra of continuous functions $C(X)$, as an ``algebraic variety",
specified by an infinite system   of quadratic equations.

Buchstaber and Rees have recently extended this to all symmetric
powers $\Sym^n(X)$ using their notion of the Frobenius
$n$-homomorphisms.

We give a simplification and a further extension of this theory,
which is based, rather  unexpectedly, on results from super linear
algebra.
\end{abstract}

\maketitle


\section{Introduction}

In 1939 Gelfand and Kolmogorov  proved~\cite{gelfand-kolmogorov}
that any compact Hausdorff topological  space $X$ is canonically
embedded into the infinite-dimensional vector space $C(X)^* $, the
dual space of the algebra of continuous functions $C(X)$, as an
``algebraic variety" specified by the infinite system   of quadratic
equations $\f (1) = 1$ and $\f(a^2)=\f(a)^2$ for linear functionals
$\f\in C(X)^*$ indexed by elements $a\in C(X)$. Recently Buchstaber
and Rees have suggested a generalization of the Gelfand--Kolmogorov
theorem based on their notion of an $n$-homomorphism or  `Frobenius
$n$-homomorphism'. They have showed that in fact all symmetric
powers $\Sym^n (X)$ of the topological space $X$ are canonically
embedded into $C(X)^* $. To this end, the quadratic equations $\f
(1) = 1$ and $\f(a^2)=\f(a)^2$, specifying the algebra
homomorphisms, have to be replaced by more complicated algebraic
equations. See~\cite{buchstaber_rees:2004, buchstaber_rees:2006} and
references therein. We have managed to find a different approach and
a further generalization for this theory~\cite{tv:gensym}, which is
motivated, rather unexpectedly, by  ideas coming from considering
linear operators acting on a superspace~\cite{tv:ber}.

In the topic of this paper we see an interaction of ideas coming
from different sources, some classical, and some quite new. They
are: Frobenius's higher group characters; the Gelfand--Kolmogorov
theorem;   supergeometry and linear algebra on superspace (Berezin);
multi-valued groups and the corresponding analog  of Hopf algebras
(Buchstaber and Rees). The latter lead
to~\cite{buchstaber_rees:2002, buchstaber_rees:2004,
buchstaber_rees:2006}.  The study of linear operators on superspace
lead to~\cite{tv:ber}.

The main question that we shall discuss may be stated as follows.
Consider a linear map between   algebras $A$ and $B$:
\begin{equation*}
    \f\co A\to B\,.
\end{equation*}
What can be said about such a map? What are `good classes' of maps
of algebras? Our algebras are associative,  with a unit, and
commutative. (This can be slightly relaxed.) We consider algebras
over $\RR$ or $\CC$.

Suppose $\f$ is an algebra homomorphism. The algebra homomorphisms
have a clear geometrical meaning. Among all statements elaborating
such an `algebraic--functional duality', let us quote the following:

\begin{thm}[Gelfand--Kolmogorov, 1939] Let $C(X)$ be the algebra of continuous functions on a compact Hausdorff
topological space $X$. Then there is a one-to-one correspondence
between the algebra homomorphisms $C(X)\to \RR$ and the points of
$X$. (All homomorphisms are  the evaluation homomorphisms at   $x\in
X$).
\end{thm}

Here the algebra $A=C(X)$ is considered purely algebraically,
without a topology. This theorem is less known than its analog where
$A$ is considered as a normed ring and homomorphisms are assumed to
be continuous.

Since the homomorphism condition $\f(ab)=\f(a)\f(b)$ can be
re-written, by using polarization, as $\f(a^2)=\f(a)^2$ for all
$a\in A$, we arrive at the system of quadratic equations in the
space $A^*$ mentioned above. These equations describe the image of
the embedding of $X$ into $A^*$. Such an interpretation has been
recently emphasized by Buchstaber and Rees, who gave an extension to
all symmetric powers $\Sym^n(X)$.

In the main text below we explain a new idea allowing to obtain  the
statement of Buchstaber and Rees very simply. Moreover, following
this path we obtain a further generalization. The main idea comes
from our recent work on Berezinians and exterior
powers~\cite{tv:ber}. In~\cite{tv:gensym} one can find a more formal
exposition. (The e-print version of~\cite{tv:gensym} contains an
appendix with details missing in the journal version.)

The paper is based on the talk by the second author at the XXVI
Workshop on Geometric Methods in Physics (Bia{\l}owie\.{z}a, Poland,
June--July 2007).

\section{A generalization of ring homomorphisms}

Motivated by their work on multi-valued groups, namely, by the
properties of the algebras of functions on such generalization of
groups, Buchstaber and Rees suggested the notion of
{$n$-homomorphisms} of algebras, where $n=1,2,3,\ldots\ $ Here
$1$-homomorphisms are ordinary algebra homomorphisms.

Recall the following construction, which can be traced back to
Frobenius. For a given linear map $\f\co A\to B$,  define maps
$\F_n\co A\times \ldots \times A\to B$ by induction:
$\F_1(a)=\f(a)$ and
\begin{multline*}\label{eq.deffrob}
    \Phi_{k+1}(a_1,\ldots,a_{k+1})=f(a_1)\Phi_k(a_2,\ldots,a_{k+1})\\
    -\Phi_k(a_1a_2,\ldots,a_{k+1})-\ldots
    -\Phi_k(a_2,\ldots,a_1a_{k+1})\,.
\end{multline*}
In Frobenius's original work this was applied to a character  of a
linear representation  of a finite group,  producing the so-called
`Frobenius higher characters'. Although the definition is not
manifestly symmetric, one can easily show by induction that the
multilinear functions $\Phi_n$ are symmetric in their arguments. It
follows that it is sufficient to consider them on the diagonal.

\begin{de} An \textit{$n$-homomorphism} $\f\co A\to B$ is a linear
map such that $\f(1)=n$ and $\F_{n+1}=0$.
\end{de}

We shall say more about properties of $n$-homomorphisms in the next
sections.

The main algebraic result of Buchstaber and Rees is the following.
\begin{thm}[Buchstaber--Rees, 2002] There is a one-to-one correspondence between the $n$-homomorphisms
$A\to B$ and the algebra homomorphisms $S^nA\to B$.
\end{thm}

Here $S^nA\subset A^{\otimes n}$ is the symmetric power of $A$
considered as a subalgebra of the tensor power $A^{\otimes n}$.
Geometrically this statement gives a canonical  embedding of the
symmetric power $\Sym^n (X)=X^n/S_n$ of a topological space $X$ into
$C(X)^*$ by a system of algebraic equations of higher order.

\begin{exa} Let $n=2$.
The embedding $\Sym^2 (X) \to C(X)^*$ is given by the formulas
\begin{equation*}
    [x_1,x_2]\mapsto \f=\ev_{[x_1,x_2]} \quad \text{where} \quad
    \ev_{[x_1,x_2]}(a)=a(x_1)+a(x_2)\,.
\end{equation*}
The equations for a linear functional $\f\co C(X)\to \RR$ are
\begin{equation*}
    \f(1)=2 \quad \text{and} \quad \begin{vmatrix}
                                     \f(a) & 1 & 0 \\
                                     \f(a^2) & \f(a) & 2 \\
                                     \f(a^3) & \f(a^2) & \f(a) \\
                                   \end{vmatrix}=0 \quad \text{for all $a\in
                                   C(X)$}\,.
\end{equation*}
(The last equation is nothing but $\Phi_3=0$.)
\end{exa}

Thus, Buchstaber and Rees introduced a class of maps of algebras
generalizing homomorphisms and discovered their beautiful geometric
properties. Buchstaber and Rees's original proofs are quite hard.
See~\cite{buchstaber_rees:2004}, where earlier works are summarized.

Now we shall explain how to obtain their results very quickly and
how they can be extended. To do so we need a digression.

\section{Digression: Berezinians and exterior powers}

Recall the following definition. For an even invertible $p|q\times
p|q$  matrix,
  $A=\left(\begin{matrix}
  A_{00} & A_{01} \\
  A_{10} & A_{11} \\
\end{matrix}\right)$, the \textit{Berezinian} or
\textit{superdeterminant} is defined by
\begin{equation*}
   \Ber A=\frac{\det \left(A_{00}-A_{01}A^{-1}_{11}A_{10}\right)}{\det
   A_{11}}\,.
\end{equation*}
It is related with the \textit{supertrace}
$\str A =\tr A_{00}-\tr A_{11}$ by Liouville's relation
\begin{equation*}
    e^{\str A}=\Ber e^A\,.
\end{equation*}

In the ordinary case $q=0$, $\Ber=\det$ and it is given by the
action on the top exterior power of a vector space. In the super
case, there is no such thing as the `top exterior power': the
sequence $\wed^k(V)$ is infinite to the right. At the first glance
there is no relation between $\Ber$ and exterior powers. However,
the following was recently discovered.

\begin{thm}[\cite{tv:ber}, 2003] If $\dim V=p|q$, then the following holds.

\textbf{(1)}  The exterior powers $\wed^k(V)$ satisfy  recurrence
relations  with $q+1$ terms  in an appropriate Grothendieck ring for
$k\geq p-q+1$. For any  linear operator $A$ on $V$ there are
`universal recurrence relations' for the  traces $\str \wed^k(A)$.
This can be expressed by the equations
\begin{equation*}
\begin{vmatrix}
      c_k  & \dots & c_{k+q}  \\
      \dots & \dots & \dots \\
      c_{k+q}  & \dots & c_{k+2q}  \\
    \end{vmatrix}=0
\end{equation*}
for $k\geq p-q+1$. Here $c_k$ are either  $\str \wed^k(A)$ or
$\wed^k V$.

\textbf{(2)} The Berezinian of a linear operator can be expressed as
a ratio of polynomial invariants:
\begin{equation*}\label{eqberezinian1}
    \Ber A=\frac{\begin{vmatrix}
      c_{p-q} & \ldots & c_p \\
      \ldots & \ldots & \ldots \\
      c_p & \ldots & c_{p+q} \\
    \end{vmatrix}}{\begin{vmatrix}
      c_{p-q+2} & \ldots & c_{p+1} \\
      \ldots & \ldots & \ldots \\
      c_{p+1} & \ldots & c_{p+q} \\
    \end{vmatrix}}=\frac{|c_{p-q}\ldots c_p|_{q+1}}{|c_{p-q+2}\ldots
    c_{p+1}|_{q}}\,,
\end{equation*}
where $c_k=\str \wed^k(A)$.
\end{thm}

(The determinants involved in formulas above are the so-called
Hankel determinants, which are   minors of the infinite `Hankel
matrix' with the entries $c_{ij}=c_{i+j}$ corresponding to an
infinite sequence $c_k$.)

The crucial tool for obtaining these and other results
in~\cite{tv:ber} is the rational \textit{characteristic function} of
a linear operator
\begin{equation*}
     R_A(z):=\Ber(1+zA)\,,
    \end{equation*}
for which we consider expansions at zero and at infinity. As we
shall see, this provides a new approach to the
Gelfand--Kolmogorov--Buchstaber--Rees theory.

\section{Characteristic function for a map of algebras}

Let $A$ and $B$ be commutative associative algebras with unit.
Consider an arbitrary linear map $\f\co A\to B$. Mimicking
constructions above, let us introduce the \textit{characteristic
function} for $\f$ as
\begin{equation*}
    R(\f,a,z):=e^{\f \ln(1+az)}\,,
\end{equation*}
where $a\in A$ and $z$ is a formal parameter. Initially $R(\f,a,z)$
is just a formal power series.

\begin{exa} Let $\f(a)=\str \rho (a)$ for a matrix representation $\rho$ of
the algebra $A$. Then $R(\f,a,z)=\Ber(1+\rho(a)z)=R_{\rho(a)}(z)$,
the characteristic function of the operator $\rho(a)$.
\end{exa}

Let us turn to general maps of algebras $\f$.

\begin{exa} If $\f$ is an algebra homomorphism, then
$R(\f,a,z)=1+\f(a)z$, i.e., a linear polynomial.
\end{exa}

We see that algebraic properties of the map $\f$ are reflected in
functional properties of $R(\f,a,z)$ w.r.t. the variable $z$. What
if   $R(\f,a,z)$ is a polynomial of degree $n$? We shall show in the
next section that this corresponds to the $n$-homomorphisms of
Buchstaber and Rees.

First let us discuss the general properties of $R(\f,a,z)$. They are
as follows (see~\cite{tv:gensym}).

$R(\f,a,z)$ satisfies the exponential property
$R(\f+\g,a,z)=R(\f,a,z)R(\g,a,z)$.

$R(\f,a,z)$ has the explicit power expansion at zero
\begin{equation*}
    R(\f,a,z)=1+\ps_1(\f,a)z+\ps_2(\f,a)z^2+\ldots
\end{equation*}
where $\ps_k(\f,a)=P_k(s_1,\ldots,s_k)$ with $s_k=\f(a^k)$ and $P_k$
being the classical Newton polynomials giving expression of
elementary symmetric functions via sums of powers:
\begin{equation*}
    P_k(s_1,\ldots,s_k)=\frac{1}{k!}\begin{vmatrix}
                          s_1    & 1          & 0      & \dots & 0 \\
                          s_2    & s_1        & 2      & \dots & 0 \\
                          \dots  & \dots      & \dots  & \dots &\dots \\
                          s_{k-1}& s_{k-2}    & s_{k-3}& \dots & k-1 \\
                          s_k    & s_{k-1}    & s_{k-2}& \dots & s_1
                        \end{vmatrix}\,.
\end{equation*}
By induction one can check that $\Phi_k(a,\ldots,a)=k!\ps_k(\f,a)$,
for the  terms of the Frobenius recursion restricted to the
diagonal.

Suppose now that $R(\f,a,z)$ extends to a genuine function of $z$
regarded, say, as a complex variable. Consider its behaviour at
infinity. By a formal transformation one can see that
$R(\f,a,z)=z^{\f(1)}e^{\f\ln a}e^{\f \ln(1+a^{-1}z^{-1})}$. In
particular, for $a=1$ we have $R(\f,1,z)=(1+z)^{\f(1)}$. Assuming
that $R(\f,a,z)$ has no essential singularity we get that
$\f(1)=\chi\in \ZZ$ is an integer, which is  the order of the pole
at infinity. Hence we have the expansion $R(\f,a,z)=\sum_{k\leq
\chi} \ps^*_k(\f,a)z^k$ at infinity, where $\ps_k^*(\f,a):=e^{\f\ln
a}\ps_{\chi-k}(\f,a^{-1})$. Denote the leading term of the expansion
\begin{equation*}
    \ber(\f,a):=e^{\f\ln a}
\end{equation*}
and call it, the \textit{$\f$-Berezinian} of $a\in A$.

One can immediately see that $\f$-Berezinian is multiplicative:
$$\ber(\f,a_1a_2)=\ber(\f,a_1)\ber(\f,a_2).$$

\section{Application:  Buchstaber--Rees theory}

Suppose that the characteristic function $R(\f,a,z)$ is a polynomial
for all $a$. In particular it follows that $\chi=\f(1)$ must be
positive; denote it $n\in\NN$. Hence it is  the degree of
$R(\f,a,z)$ for all $a$. So $\ps_k(\f,a)=0$ for all $k\geq n+1$ and
all $a\in A$. This is equivalent to the equations $\f(1)=n\in \NN$
and $\F_{n+1}(\f,a_1,\ldots,a_{n+1})=0$ for all $a_i$, which is
precisely the definition of an $n$-homomorphism according to
Buchstaber and Rees~\cite{buchstaber_rees:2004}.

Various properties of $n$-homomorphisms immediately follow from this
description. For example, the exponential property of the
characteristic function implies that the sum of an $n$-homomorphism
and an $m$-homomorphism is an $(n+m)$-homomorphism. Similarly one
can deduce that the composition of  $n$-homomorphism and an
$m$-homomorphism is an $nm$-homomorphism. (These results were
originally obtained much harder, see~\cite{buchstaber_rees:2004,
buchstaber_rees:2006}.)

The main theorem of Buchstaber and Rees can   be easily obtained as
follows.

Since $\f$-Berezinian is  multiplicative, and for $n$-homomorphisms
$\ber(\f,a)=\ps_n(\f,a)$,  the  function $\ps_n(\f,a)$ is
multiplicative in $a$. Therefore its polarization
$\frac{1}{n!}\F_n(\f,a_1,\ldots,a_n)$ yields an algebra homomorphism
$S^nA\to B$. Thus a one-to-one correspondence between the
$n$-homomorphisms $A\to B$ and the algebra homomorphisms $S^nA\to B$
is established. The transparency of this proof illustrates the power
of our approach. (The multiplicativity of
$\frac{1}{n!}\F_n(\f,a,\ldots,a)$ was the hardest part of the
original proof~\cite{buchstaber_rees:2002}; it was obtained there by
using a non-trivial combinatorics.)

\section{Further extension: generalized symmetric powers}

Suppose the characteristic function $R(\f,a,z)$ is not a polynomial,
but a rational function. We arrive at a further generalization of
ring homomorphisms.

\begin{de} We call a linear map $\f\co A\to B$ a
\textit{$p|q$-homomorphism} if   $R(\f,a,z)$ can be written as the
ratio of polynomials of degrees $p$ and $q$.
\end{de}

We have $\chi=\f(1)=p-q$ for $p|q$-homomorphisms.

\begin{exas} The negative $-\f$ of a ring homomorphism
$\f$ is a $0|1$-homomorphism.  The difference $\f_{(p)}-\f_{(q)}$ of
a $p$-homomorphism $\f_{(p)}$ and a $q$-homomorphism $\f_{(q)}$ is a
$p|q$-homomorphism. In particular, a linear combination of algebra
homomorphisms of the form $\sum n_{\a} \f_{\a}$ where $n_{\a}\in
\ZZ$ is a $p|q$-homomorphism with $\chi=\sum n_{\a}$,
$p=\sum\limits_{n_{\a}>0} n_{\a}$, and $q=-\sum\limits_{n_{\a}<0}
n_{\a}$.  It all follows from the exponential property of the
characteristic function.
\end{exas}

By using formulas from~\cite{tv:ber}, the condition that $\f\co A\to
B$ is a $p|q$-homomorphism can be expressed by the equations
\begin{equation}\label{eq.vanhank}
\f(1)=p-q \quad \text{and} \quad
\begin{vmatrix}
      \ps_k(\f,a)  & \dots & \ps_{k+q}(\f,a)  \\
      \dots & \dots & \dots \\
      \ps_{k+q}(\f,a)  & \dots & \ps_{k+2q}(\f,a)  \\
    \end{vmatrix}=0
\end{equation}
(the Hankel determinant), for all $k\geq p-q+1$.

What is the geometrical meaning of this notion?

Consider a  topological space $X$. We define its  \textit{$p|q$-th
symmetric power} $\Sym^{p|q}(X)$  as the identification space of
$X^{p+q}$ with respect to the action of $S_p\times S_q$ and the
relations
$$(x_1,\ldots,x_{p-1},y,x_{p+1}\ldots,x_{p+q-1},y)\sim
(x_1,\ldots,x_{p-1},z,x_{p+1}\ldots,x_{p+q-1},z)\,.
$$
The algebraic analog of $\Sym^{p|q}(X)$ is the \textit{$p|q$-th
symmetric power} $S^{p|q}A$ of a commutative associative algebra
with unit $A$. We define $S^{p|q}A$ as the subalgebra
$\mu^{-1}\left(S^{p-1}A\otimes S^{q-1}A\right)$ in $S^pA\otimes
S^qA$ where $\mu\co S^pA\otimes S^qA \to S^{p-1}A\otimes
S^{q-1}A\otimes A$ is the multiplication of the last arguments.

\begin{exa} For $A=\CC[x]$, the algebra $S^{p|q}A$ is the algebra of
all polynomial invariants of $p|q$ by $p|q$ matrices.  (This is a
non-trivial statement, see in~\cite{tv:ber}.)
\end{exa}

There is a relation between algebra homomorphisms $S^{p|q}A\to B$
and $p|q$-homomorphisms $A\to B$. To each homomorphism $S^{p|q}A\to
B$ canonically corresponds a $p|q$-homomorphism $A\to B$.

\begin{exa}
An element $\xp=[x_1,\ldots,x_{p+q}]\in \Sym^{p|q}(X)$ defines the
$p|q$-homomorphism   $\ev_{\xp}\co C(X)\to \RR$:
\begin{equation*}
    a\mapsto a(x_1)+\ldots +a(x_{p})-\ldots
 -a(x_{p+q})\,.
\end{equation*}
This gives a natural map $\Sym^{p|q}(X)\to A^*$, where $A=C(X)$,
which generalizes the Gelfand--Kolmogorov and Buchstaber--Rees maps
(in fact, an embedding). The image of $\Sym^{p|q}(X)$ in $A^*$
satisfies equations~\eqref{eq.vanhank} where $\f=\ev_{\xp}$. It is a
system of polynomial equations for `coordinates' of a linear map
$\f\in A^*$.
\end{exa}

A conjectured statement is that the solutions of
equations~\eqref{eq.vanhank} give precisely the image of
$\Sym^{p|q}(X)$. This would be an exact analog of the
Gelfand--Kolmogorov and Buchstaber--Rees theorems. The corresponding
algebraic statement would be a one-to-one correspondence between the
$p|q$-homomorphisms $A\to B$ and the algebra homomorphisms
$S^{p|q}A\to B$.


\begin{theacknowledgments}
  It is a pleasure to thank the organizers of the annual Workshops
  on Geometric Methods in Physics in Bia{\l}owie\.{z}a and  particularly Prof. Anatole
  Odzijewicz for the hospitality and the exceptionally inspiring atmosphere at the meetings.
\end{theacknowledgments}

\end{document}